\documentstyle[12pt]{article}
\begin{document}
\textheight 22 cm
\topmargin -30.0 mm
{\hfill  TECHNION--PH-95-7}

\hspace{10.0cm}   May, 1995

\vspace{6.0cm}
\centerline{\Large On the formation of black holes}
\centerline{\Large in non-symmetric gravity}
\vspace{1.0cm}
\centerline{Lior M. Burko and Amos Ori, Department of Physics}
\centerline{Technion -- Israel Institute of Technology,
32000 Haifa, Israel}
\vspace{2.5cm}
\centerline{Abstract}
It has been recently suggested that
the Non-symmetric Gravitational Theory
(NGT) is free of black holes. Here,
we study the linear version of NGT. We find that even
with spherical symmetry the skew part of
the metric is generally
non-static. In addition,
if the skew field is initially regular, it will
remain regular everywhere and,
in particular, at the horizon. Therefore, in
the fully-nonlinear theory,
if the initial skew-field is sufficiently
small, the formation of a black hole is to be anticipated.

\vspace{1.0cm}
 \vspace{0.5cm} 04.70.Bw
\newpage
\textheight 22 cm
\topmargin -30.0 mm

General Relativity (GR) is a theory of gravitation, in which gravity is
manifested by the curvature of spacetime, which is described by Riemannian
geometry \cite{einstein}.
Field theories which use non-Riemannian geometry have been formulated by
Einstein \cite{einstein2}, Schr\"{o}dinger \cite{schrodinger},
Einstein and Straus \cite{einsteinandstraus}, and more recently by Moffat
\cite{moffat,moffat1} and Klotz \cite{klotz}.
In non-Riemannian
geometry the metric tensor $g_{\mu\nu}$ is not assumed to be symmetrical in
its two indices. This property complicates the geometry considerably, and
induces torsion in spacetime. Einstein \cite{einstein2} formulated
a non-symmetric field
theory as part of his quest for a Unified Field Theory,
namely, a unified theory of classical gravity and electromagnetism.

Recently, there has been growing interest in Non-symmetric Gravitation
Theory (NGT) for motivations different than Einstein's.
Cornish and Moffat (CM) \cite{cornishandmoffat,cornishandmoffat1}
studied a class of
exact static spherically-symmetric solutions to the NGT field
equations. This class depends on the two parameters $m$ and $s$, where $m$
is the source's mass, and $s$ determines the strength of the
skew part of the metric tensor (that is, at large distance,
$r\gg m$,
this skew part is proportional to $s$). In all these solutions,
there are no trapped surfaces, and consequently there are no black holes.
Based on these static solutions, CM suggested that NGT was free of black
holes (and, thereof, of spacetime singularities)
\cite{cornishandmoffat,cornishandmoffat1,science}.

It is remarkable that even for arbitrarily small $s$, the
static skew field ``destroys" the horizon. That is, even if the skew
field ({\it i.e.},
the skew part of the metric tensor) is arbitrarily
small at $r\gg m$, in the static solution
it grows in an uncontrolled way on the approach to
$r=2m$  --  until it becomes so strong that it modifies the geometry
dramatically and prevents the formation of trapped surfaces.
This behavior is nicely demonstrated in the context of linearized NGT. In
linearized NGT, the skew field is regarded as an infinitesimally-small
perturbation over the standard, symmetric metric. The linear analogue of
the model analyzed by CM is that of a static, spherically-symmetric,
linearized skew field on a Schwarzschild background.
One then finds (see below)
that the linearized
skew field diverges at $r=2m$. This linear divergence indicates the
effectiveness of the skew field, and its
ability to ``destroy" the black hole, in the context of fully non-linear
NGT.

Of course, before making any definitive statements about the
existence or non-existence of black holes in NGT, one must address the
following question: Is the above mentioned phenomenon (the absence of
a black hole in the static spherically-symmetric solutions) a generic
characteristic of NGT, or a result of the symmetry (staticity)
imposed? Answering this question requires an investigation of the
nature of the generic dynamical NGT solutions. This is
an extremely hard task, because NGT is much more
complicated than GR (and, of course, its general dynamic solution is as yet
unknown). Fortunately, it is possible to translate the above question to
the context of linearized NGT: Is the divergence of the linearized skew
field at the Schwarzschild radius a generic feature of linearized NGT, or a
consequence of the assumption of staticity (of the skew field)? If the
generic linearized solution were divergent at the background's horizon,
an important dynamical effect would be anticipated in the
fully-nonlinear NGT. On the other hand, if the linearized skew field is
found to be generically regular at the horizon, the situation is different:
Then (at least for sufficiently-small initial skew field), the linearized
solution is likely to be a good approximation to the full theory, and no
drastic effects are expected to occur at the horizon. In such a case, we
should expect a gravitational collapse to proceed pretty much like in GR
-- and, in particular, a formation of a black hole is to be anticipated.

The goal of this Letter is to address the above question. We shall
study the dynamical behavior of a linearized skew field on a GR
background and apply this formalism to spherically-symmetric skew field on
Schwarzschild. We shall show that the linearized equation possesses a
well-posed initial-value formulation. An important result is that, even in
spherical symmetry, the skew field need not be static. Moreover, for
regular initial data on some spacelike hypersurface $\Sigma$, no divergence
occurs anywhere in the entire domain of dependence. In particular, the
dynamics of the skew field at the horizon is perfectly regular. Our
conclusion is, therefore, that if the initial skew field is sufficiently
small, a black hole is likely to form in gravitational collapse -- just
like in standard GR.

The vacuum field equations of NGT are:
\begin{eqnarray}
g_{\mu\nu ,\sigma}-g_{\rho\nu}\Gamma^{\rho}_{\mu\sigma}
-g_{\mu\rho}\Gamma^{\rho}_{\sigma\nu}=0,
 \label{metricity} \\
\left(\sqrt{-g}g^{[\mu\nu]}\right)_{, \nu}=0,
 \label{gauge}\\
R_{(\alpha\beta)}=0,
 \label{vacuum}\\
R_{[\alpha\beta],\gamma}+R_{[\beta\gamma],\alpha}+R_{[\gamma\alpha],\beta}=0
, \label{cyclic}
\end{eqnarray}
where $g_{\mu\nu}$ is the non-symmetric metric tensor, $g$ is
its determinant,
$R_{\alpha\beta}$ is the generalized Ricci tensor [see Eq. (\ref{ricci})
below],
and $\Gamma^{\alpha}_{\beta\gamma}$ is the non-symmetric affine
connection. The inverse metric $g^{\mu\nu}$ is defined by
$g^{\mu\nu}g_{\mu\sigma}=g^{\nu\mu}g_{\sigma\mu}=\delta^{\nu}
_{\;\sigma}$.

We now consider the linearized NGT. Namely, we assume that the skew
part of the metric tensor, $h_{\mu\nu}$,
is a small perturbation over the symmetric GR
metric, and develop the field equation
to first order in this perturbation.
Denoting all background fields by an overhat, we write $g_{\mu\nu}
\equiv \hat{g}_{\mu\nu}+h_{\mu\nu}$, $\Gamma^{\rho}_{\mu\nu}
\equiv \hat{\Gamma}^{\rho}_{\mu\nu}+D^{\rho}_{\mu\nu}$, and
$R_{\alpha \beta}\equiv \hat{R}_{\alpha\beta}+Q_{\alpha\beta}$. Here,
$\hat{g}_{\mu\nu}$ is a standard, symmetric GR metric, and
$\hat{\Gamma}^{\rho}_{\mu\nu}$ and $\hat{R}_{\alpha\beta}$ are the standard
connection and Ricci tensor, respectively, associated with this background
metric.  Note
the symmetry features of the various entities: By definition, we have
$\hat{g}_{(\mu\nu )}=\hat{g}_{\mu\nu}$,
$\hat{\Gamma}^{\rho}_{(\mu\nu )}=\hat{\Gamma}^{\rho}_{\mu\nu}$,
$\hat{R}_{(\alpha\beta )}=\hat{R}_{\alpha\beta }$, and $h_{[\mu\nu ]}
=h_{\mu\nu}$. We shall show below that $D^{\alpha}_{[\beta\gamma ]}=
D^{\alpha}_{\beta\gamma}$ and $Q_{[\alpha\beta ]}=Q_{\alpha\beta}$.
The background metric $\hat{g}_{\mu\nu}$ is taken to be
vacuum, {\it i.e.}, $\hat{R}_{\mu\nu}=0$.

{}From the metricity equation (\ref{metricity})
we find, to the linear order in the skew field, that
\begin{eqnarray}
\hat{g}_{\rho\beta}D^{\rho}_{\alpha\gamma}+
\hat{g}_{\alpha\rho}D^{\rho}_{\gamma\beta}=h_{\alpha\beta ;\gamma},
 \label{linearized metricity}
\end{eqnarray}
where a semicolon denotes covariant differentiation with respect to the
GR background $\hat{g}_{\mu\nu}$.
Solving Eq. (\ref{linearized metricity}) we find that
\begin{eqnarray}
D^{\alpha}_{\beta\gamma}=
D^{\alpha}_{[\beta\gamma]}=\frac{1}{2}\hat{g}^{\alpha\delta}
\left(h_{\delta\gamma ;\beta}+h_{\beta\delta ;\gamma}
+h_{\beta\gamma ;\delta}\right).
\end{eqnarray}
Next, we linearize Eq. (\ref{gauge}). To linear order we have
$g^{[\mu\nu ]}=-h^{\mu\nu}$. (We use the background metric
$\hat{g}_{\alpha\beta}$ to raise or lower indices.) Eq. (\ref{gauge}) is
thus reduced to
\begin{eqnarray}
h^{\alpha\beta}_{\;\;\; ;\beta}=h^{\beta\alpha}_{\;\;\; ;\beta}=0.
 \label{Z1}
\end{eqnarray}

The generalized (Hermitianized) Ricci tensor is defined in NGT by
\cite{einstein2}
\begin{eqnarray}
R_{\alpha\beta}=\Gamma^{\rho}_{\alpha\beta ,\rho}-\frac{1}{2}
\left(\Gamma^{\rho}_{(\alpha\rho),\beta}+\Gamma^{\rho}_{(\rho\beta),\alpha}
\right)-\Gamma^{\rho}_{\alpha\sigma}\Gamma^{\sigma}_{\rho\beta}
+\Gamma^{\rho}_{\alpha\beta}\Gamma^{\sigma}_{(\rho\sigma)}.
 \label{ricci}
\end{eqnarray}
Expanding this equation to the first order in the perturbation,
we find that
\begin{eqnarray}
Q_{\alpha\beta}=Q_{[\alpha\beta]}=D^{\rho}_{\alpha\beta ;\rho},
\end{eqnarray}
or, equivalently,
\begin{eqnarray}
Q_{\beta\gamma}=\frac{1}{2}\hat{g}^{\alpha\delta}\left(
h_{\delta\gamma ;\beta\alpha}+h_{\beta\delta ;\gamma\alpha}
+h_{\beta\gamma ;\delta\alpha}\right).
 \label{cov}
\end{eqnarray}

Recalling the non-commutivity of covariant derivatives,
we re-write Eq. (\ref{cov}) as
\begin{eqnarray}
Q_{\beta\gamma}=\frac{1}{2}\hat{g}^{\delta\alpha}\left(
h_{\delta\rho}\hat{R}^{\rho}_{\;\gamma\beta\alpha}
+h_{\rho\gamma}\hat{R}^{\rho}_{\;\delta\beta\alpha}+
h_{\beta\rho}\hat{R}^{\rho}_{\;\delta\gamma\alpha}+
h_{\rho\delta}\hat{R}^{\rho}_{\;\beta\gamma\alpha}\right) +\nonumber \\
\frac{1}{2}\left( h^{\alpha}_{\;\gamma ;\alpha\beta}
+h^{\;\alpha}_{\beta\; ;\alpha\gamma} +\hat{g}^{\delta\alpha}
h_{\beta\gamma ;\delta\alpha}\right).
 \label{Q}
\end{eqnarray}
where $\hat{R}^{\rho}_{\;\gamma\beta\alpha}$ is the
background Riemann curvature tensor.
In view of $\hat{R}_{\alpha\beta}=0$ and Eq. (\ref{Z1}),
Eq. (\ref{Q}) becomes
\begin{eqnarray}
Q_{\beta\gamma}=\frac{1}{2}\hat{g}^{\delta\alpha}h_{\beta\gamma
;\delta\alpha}+2\hat{g}^{\delta\alpha}h_{\delta\rho}
\hat{R}^{\rho}_{\; \alpha\beta\gamma}.
 \label{Q2}
\end{eqnarray}
Linearizing Eqs. (\ref{vacuum}) and (\ref{cyclic}), one finds that
the former is automatically satisfied by $Q_{\beta\gamma}$,
and Eq. (\ref{cyclic}) reduces to
\begin{eqnarray}
Q_{\alpha\beta ,\gamma}+Q_{\gamma\alpha ,\beta}+Q_{\beta\gamma ,\alpha}=0.
 \label{Z2}
\end{eqnarray}
Equation (\ref{Z2}) [with the identity (\ref{Q2})],
together with the constraint (\ref{Z1}), are the linearized
vacuum NGT equations for $h_{\mu\nu}$.

Our analysis so far was quite generic: We did not make any assumptions
about any symmetry of either $\hat{g}_{\mu \nu }$ or $h_{\mu \nu }$.
We shall now restrict
attention to spherical symmetry. Namely, we shall take
$\hat g_{\mu \nu }$ to be Schwarzschild, and $h_{\mu \nu }$ to be
spherically-symmetric.
We start from the spherically-symmetric metric used by
CM (see, {\it e.g.}, Eq. (5) in Ref. \cite{cornishandmoffat}),
and allow the three non-trivial metric
functions -- namely,
$\alpha$ , $\gamma$ and $f$ -- to depend on both $r$ and $t$:
\begin{eqnarray}
g_{\mu\nu}=\left(\begin{array}{cccc}
\gamma (r,t)& 0 & 0 & 0 \\
0 & -\alpha (r,t)& 0 & 0 \\
0 & 0 & -r^{2} & f(r,t)\sin\theta \\
0 & 0 & -f(r,t)\sin\theta & -r^{2}\sin^{2}\theta
\end{array}\right) .
 \label{metric}
\end{eqnarray}
(The most general spherically-symmetric metric
may also include a nonzero metric function
$g_{[rt]}$ \cite{papapetrou}.
Here, we follow CM and restrict attention to the simpler case,
$g_{[rt]}=0$.)
We now linearize the field equations in $f$. The zeroth-order equation
$\hat{R}_{\mu \nu }=0$ immediately implies that the background metric is
the Schwarzschild solution:
$\gamma =1/\alpha =1-2m/r$, so we only need to calculate $f$.
Equation (\ref{Z1}) is automatically satisfied by the skew part of
(\ref{metric}), and
we only need to consider Eq. (\ref{Z2}). A straightforward calculation,
based on Eq. (\ref{Q2}), yields that the only
non-vanishing components of $Q_{\mu \nu }$ are
\begin{eqnarray}
Q_{\theta\phi}=-Q_{\phi\theta}=\left[
\frac{1}{2}\left(\frac{\ddot{f}}{\gamma}-\frac{f''}
{\alpha}\right)+\frac{f'}{\alpha r}+\frac{1}{2}\frac{f'\alpha '}
{\alpha^{2}}-2\frac{f\alpha '}{\alpha^{2}r}\right]\sin\theta,
\end{eqnarray}
where a dot and a prime denote partial
differentiation with respect to $t$ and  $r$, correspondingly.
[We have also derived this equation directly, by calculating $R_{\mu \nu }$
from the (time-dependent) metric
(\ref{metric}) in the fully nonlinear NGT,
and then linearizing it in $f$.]
{}From Eq. (\ref{Z2}) it is obvious that $Q_{\theta\phi}$ cannot depend on
$r$ or $t$. The most general solution of this equation is, therefore,
$Q_{\theta\phi}=-c \sin\theta$,
where $c$ is some real constant (see also Ref. \cite{vanstone}).
It can be shown, however, that for $c\ne 0$ the spacetime is not
asymptotically-Minkowski \cite{cornish1}. We shall therefore focus
attention here on the case $c=0$.
The field equation for $f$ will thus be
\begin{eqnarray}
\frac{1}{2}\left(\frac{\ddot{f}}{\gamma}-\frac{f''}
{\alpha}\right)+\frac{f'}{\alpha r}+\frac{1}{2}\frac{f'\alpha '}
{\alpha^{2}}-2\frac{f\alpha '}{\alpha^{2}r}=0.
 \label{f}
\end{eqnarray}
In the static limit, {\it i.e.}, when $\dot{f}$ is taken to vanish,
we recover from Eq. (\ref{f}) the linear analogue of the CM equation
for $f$ (see, in particular,  Eq. (2.4) of Ref. \cite{cornish}).
One can easily verify that, in the static limit, the linearized $f$
diverges logarithmically at $r=2m$. This is just the linear analogue
of the behavior found by CM. Here, however, we are in a position to study
the dynamical content of the theory.

Equation (\ref{f}) is a linear, second-order, hyperbolic,
partial differential equation,
and consequently it possesses a well-posed initial-value formulation.
Thus, given $f$ and $\dot{f}$
on some spacelike surface, standard
theorems guarantee the existence and uniqueness of a regular
solution $f(r,t)$ throughout the domain of dependence (or, more precisely,
as
long as the background metric tensor is regular). This, by itself,
proves that $f$ does not satisfy a generalized Birkhoff's theorem
\cite{clayton}.
Namely, despite the spherical symmetry,  $f$ is generically dynamic
(for, one is allowed to chose nonzero initial $\dot{f}$).

The next stage of our analysis is to study the behavior of $f$ at the horizon.
The Schwarzschild co-ordinates are unsuitable for that purpose as they
go singular at $r=2m$. We therefore need to transform to some other
spherical co-ordinates ({\it e.g.}, Kruskal-Szekeres \cite{kruskal}). This
 transformation
is most easily done by expressing Eq. (\ref{f}) in a covariant form.
Defining a new function $k(r,t)\equiv f(r,t)/r^{2}$, one readily finds that
Eq. (\ref{f}) reduces to
\begin{eqnarray}
\hat{g}^{\mu\nu}k_{; \mu\nu}+\frac{2}{r^{2}}k=0.
 \label{k}
\end{eqnarray}

Take now any co-ordinates that cover the Schwarzschild manifold
(such as Kruskal-Szekeres), and re-express Eq. (\ref{k}) in terms of
partial derivatives. The resultant equation is obviously
a linear, second-order, hyperbolic, partial differential equation --
throughout the spacetime (with coefficients which are
regular everywhere).
Therefore, for any partial Cauchy surface $\Sigma$
in the analytically-extended Schwarzschild spacetime,
and for any choice
of regular $k$ (or $f$) and its time-derivative on it,
the existence and uniqueness of a regular
solution $k(r,t)$ [or $f(r,t)$] throughout $D^+(\Sigma )$ is guaranteed.
In particular, $f$ is regular at the horizon.

We have found that if the linearized skew function $f$ is initially
regular, it will remain regular throughout the domain of dependence (except
possibly at $r=0$)
and, in particular, at the event horizon. Note that there
is no conflict between this result and the divergence of the static
linearized skew field at $r=2m$. From the initial-value point of view, the
linearized
static solution fails to be regular at  $r=2m$ simply because it evolved
from singular initial data. (That is, in view of the staticity, the
divergence at $r=2m$ must have been existed already on the initial
slice.) For any regular initial data, however, the skew field will remain
regular at the horizon.

Let us now discuss the implication of the above results to
nonlinear NGT. Generally, one expects a linear perturbation analysis to be
a good approximation to the original nonlinear theory as long as the
perturbation is small. If, however, the linearized perturbation
develops a divergence at some point, this may break the validity of the
linear approximation. Indeed, the divergence of the static linearized
skew field at the horizon indicates strong nonlinear effects, which
completely modify the GR geometry (at $r\le 2m$). We have found,
however, that if the initial data for the linear case
are regular, no divergence will occur.
We therefore arrive at the
following conclusion regarding the behavior of the fully-nonlinear system:
If the skew function $f$ and its time derivative
are regular and sufficiently small at the initial moment, they are likely
to remain small, and dynamically-unimportant, in the neighborhood of
$r=2m$. In particular, a black hole is expected to form -- pretty much
like in GR.\footnotemark
\footnotetext{Important dynamical effects are possible, however, near
$r=0$.} Again, there is no conflict between this result and the strong
nonlinear effect found by CM in the static case, because in the latter the
initial skew field is necessarily strong near $r=2m$.

Strictly speaking, the above considerations are restricted to the vacuum
case, {\it i.e.}, to the analytically-extended Schwarzschild spacetime.
One may therefore be concerned about the validity of our conclusion to the
situation of gravitational collapse (in spherically-symmetric gravitational
collapse matter must always be involved). The present authors regard this
as a technical difficulty, rather than an inherent one. Although our
regularity arguments are not strictly valid in the presence of matter, in
view of the above analysis there is no positive indication whatsoever for
any anomalous behavior of the skew field at the horizon (given regular and
sufficiently small initial data).

In addition, let us
imagine a non-spherical
GR background $\hat{g}_{\alpha\beta}$ describing a
dynamical gravitational collapse of pure gravitational radiation (which in
GR produces a black hole \cite{abrahams}). Consider now a small (linearized)
skew perturbation $h_{\alpha\beta}$ over this background. (We assume that
the initial data for the skew field are given on an initial hypersurface
prior to the formation of the black hole.)
The vacuum field equations are certainly valid
in that case. Although our above initial-value analysis is restricted to
spherical symmetry, it is possible to extend it to
the generic (non-spherical) case. \cite{burko} This general analysis
is beyond the scope of the present Letter, so we shall just outline it
briefly. In the generic case, one can introduce a ``vector-potential''
$A_{\mu} $ ($A_{\mu} $ is closely related to the vector $W_{\mu}$ of Ref.
\cite{moffat}), such that
$Q_{\mu \nu }=A_{[\mu ,\nu] }$. [This automatically
solves Eq. (\ref{Z2}).]
Using the Lorentz gauge,
$A^{\mu}_{\;\; ;\mu}=0$,
one can derive a system of second-order linear hyperbolic differential
equations for $A_{\mu} $ and $h_{\mu \nu }$, which is
consistent with the constraint equations [{\it i.e.,} with
Eq. (\ref{Z1}) and
$A^{\mu}_{\;\; ;\mu}=0$]. Doing so, we again obtain a well-posed
intial-value
formulation for the generic evolution of the linearized non-symmetric
field. One can now repeat the above arguments and arrive at
a similar conclusion -- this time, applied to the
formation of a non-spherical
black hole by the collapse of pure gravitational radiation: If
the initial skew field is sufficiently small, no important dynamical
effects are expected to occur on the approach to the event (or apparent)
horizon. Therefore, a black hole is expected to form, as in GR.

If, indeed, a black hole forms in NGT, what would then be its final state?
The equation satisfied by $k$  [Eq. (\ref{k})] is nothing but
the radial equation for the $l=1$ mode of a
massless scalar field. Consequently, from the analysis of Price
\cite{price}, an external observer will witness an inverse power-law decay
(in the external time $t$)
of the skew field, with a usual GR black hole as the final
state.\footnotemark\footnotetext{It is
interesting to note that
in the more general case, $c\ne 0$, a permanent skew hair will
remain after perturbations die out: Defining $\tilde{k}\equiv k+c$, one
readily finds that Eq. (\ref{k}) is recovered with $k$ replaced by $\tilde{k}$.
This means that $\tilde{k}$ will decay asymptotically to zero, and
consequently $k$ will approach $-c$.
The nature of this hair, and the implications it
has on the features of the black hole, await further investigation.
Recall, however, that this case is not asymptotically-Minkowski.}

We thank Mike Clayton, Neil Cornish and John Moffat for stimulating
discussions and helpful comments.
This research was supported in part by The Israel Science
Foundation administrated by the Israel Academy of Sciences and
Humanities.


\begin{thebibliography}{99}

\bibitem{einstein} A. Einstein, Preuss. Akad. Wiss. Berlin, Sitzber.,
778-786 (1915); {\it ibid.} 799-801 (1915).

\bibitem{einstein2} A. Einstein, {\it The Meaning of Relativity},
fifth edition
(Princeton University Press, Princeton, New-Jersey, 1956).

\bibitem{schrodinger} E. Schr\"{o}dinger, Proc. Roy. Irish Academy A
{\bf 51}, 163 (1947).

\bibitem{einsteinandstraus} A. Einstein and E.G. Straus, Annals of
Mathematics {\bf 47}, 731-741 (1946).

\bibitem{moffat} J.W. Moffat, Phys. Rev. D {\bf 19}, 3354-3358 (1979).

\bibitem{moffat1} J.W. Moffat, ``Nonsymmetric Gravitational Theory,''
University of Toronto Preprint UTPT-94-28, gr-qc/9411006 (to be published);
``Regularity Theorems in the Nonsymmetric Gravitational Theory,''
University of Toronto Preprint UTPT-95-05,
gr-qc/9504009 (to be published).

\bibitem{klotz} A.H. Klotz, {\it Macrophysics and geometry}
(Cambridge University Press, Cambridge, 1982) and references cited
therein.

\bibitem{cornishandmoffat} N.J. Cornish and J.W. Moffat, Phys. Lett. B
{\bf 336}, 337-342 (1994).

\bibitem{cornishandmoffat1} N.J. Cornish and J.W. Moffat, J. Math. Phys.
{\bf 35}, 6628-6643 (1994).

\bibitem{science} See also: F. Flam, Science {\bf 266}, 1945 (1994).


\bibitem{papapetrou} A. Papapetrou, Proc. Roy. Irish Academy A {\bf 52},
69-86 (1948).

\bibitem{vanstone} J.R. Vanstone, Can. J. Math. {\bf 14}, 568-576 (1962).

\bibitem{cornish1} N.J. Cornish, Private communication (1995).

\bibitem{cornish} N.J. Cornish, ``The Unholey Solution to Black Hole
Information Loss,'' gr-qc/9503034 (unpublished).

\bibitem{clayton} This has been shown independently by Clayton and
by Cornish, Moffat and Tatarski, although
through a completely different procedure. See
M.A. Clayton, ``Massive NGT and Spherically Symmetric
Systems,'' University of Toronto preprint UTPT-95-09,
gr-qc/9505005 (unpublished); N. J. Cornish, J. W. Moffat
and D. Tatarski, Gen. Rel. Grav., to appear (1995).

\bibitem{kruskal} M.D. Kruskal, Phys. Rev. {\bf 119}, 1743-1745 (1960);
G. Szekeres, Publ. Mat. Debrecen {\bf 7}, 285-301 (1960).


\bibitem{abrahams} A.M. Abrahams and C.R. Evans, Phys. Rev. D. {\bf 46},
R4117-4121 (1992).

\bibitem{burko} L.M. Burko, ``On the collapse of pure
gravitational radiation
in non-symmetric gravity,'' Technion Report
TECHNION-PH-95-15 (unpublished).


\bibitem{price} R.H. Price, Phys. Rev. D {\bf 5}, 2419-2438 (1972).
\end{thebibliography}
\end{document}